# Optical nanofiber integrated into an optical tweezers for particle manipulation and *in-situ* fiber probing

Ivan Gusachenko[a], Mary. C. Frawley[a,b], Viet. G. Truong[a], Síle Nic Chormaic[a*]

[a]Light-Matter Interactions Unit, OIST Graduate University, 1919-1 Tancha, Onna-son, Okinawa, Japan 9040495; [b]Physics Department, University College Cork, Cork, Ireland

## ABSTRACT

Precise control of particle positioning is desirable in many optical propulsion and sorting applications. Here, we develop an integrated platform for particle manipulation consisting of a combined optical nanofiber and optical tweezers system. Individual silica microspheres were introduced to the nanofiber at arbitrary points using the optical tweezers, thereby producing pronounced dips in the fiber transmission. We show that such consistent and reversible transmission modulations depend on both particle and fiber diameter, and may be used as a reference point for *in-situ* nanofiber or particle size measurement. Therefore we combine SEM size measurements with nanofiber transmission data to provide calibration for particle-based fiber assessment. We also demonstrate how the optical tweezers can be used to create a 'particle jet' to feed a supply of microspheres to the nanofiber surface, forming a particle conveyor belt. This integrated optical platform provides a method for selective evanescent field manipulation of micron-sized particles and facilitates studies of optical binding and light-particle interaction dynamics.

## 1. INTRODUCTION

A decade after its first realization, the laser was used to move micron-sized dielectric particles by Ashkin.[1] Nowadays, light-mediated trapping boasts a myriad of experimental applications, including cell manipulation,[2] force measurement,[3] holographic trapping,[4] angular momentum transfer,[5] and the use of optical fibers as trapping sources.[6] Recent advances include the study of light-matter coupling in optically bound structures,[7] and lab-on-a-chip integrated techniques.[8] A sister branch of optical manipulation that is subject to increasing attention exploits the potential of surface evanescent waves. Prism surfaces,[9] microscope objectives,[10] and waveguides[11] have been used as evanescent field interfaces to probe the dynamics of optical trapping and binding, including extensive studies of counter propagating and standing-wave field effects.[12]

A special case of evanescent field geometry is found in the optical nanofiber. With a diameter comparable to the wavelength of light guided within, such ultrathin optical fibers have intense evanescent fields which penetrate into the surrounding medium.[13] Being relatively easy to fabricate and integrate with other optical components, nanofibers have emerged as compact, versatile devices with a broad range of applications,[14] such as cold-atom manipulation, molecular spectroscopy,[15] and microresonator coupling.[16] Optical manipulation using nanofibers is now a field of increasing potential;[17] the evanescent field around the waist of the nanofiber is used to optically trap and propel micron-sized particles in suspension. In a manner reminiscent of conventional optical tweezers, the gradient of the evanescent field attracts nearby particles to the fiber surface. These are then propelled along the direction of light propagation via radiation pressure. Since their establishment as optical propulsion tools,[17] nanofibers have been used for bidirectional particle conveyance,[18] wavelength selective particle sorting,[19] and mass biological particle migration under photophoresis.[20] Such methods have exciting applications as particle 'conveyor belts' and sorting mechanisms in enclosed systems, particularly as their mm-scale lengths also facilitates continuous and long range trapping at any point in a sample, beyond limits achievable with conventional focused-beam tweezers.

In most cases, nanofibers are immersed in a 'particle bath' – a relatively high-density solution which allows many particles to simultaneously interact with the fiber. However, particles moving in and out of the evanescent field can cause major scattering-induced system fluctuations, making it difficult to determine the evanescent field incident on particles within a given visual frame. Microfluidic insertion is one way to address this problem.[21] Although such systems are highly relevant for mass particle sorting and filtration, they can be complex to arrange and have limitations in terms of particle/site selection and system reversibility.

*sile.nicchormaic@oist.jp; https://groups.oist.jp/light

Here, we demonstrate an integrated platform for particle manipulation using a combined optical nanofiber and optical tweezers system. This allows particles to be selectively trapped, individually or in arrays, and site-specifically introduced to the nanofiber surface. Previously we showed that addressing nanofibers in a more structured and site-specific way offers many advantages, including facilitating the study of inter-particle and particle-evanescent field interactions[22]. In this paper we demonstrate that a single silica particle of known diameter can serve as a probe to sense the local fiber diameter by observing the nanofiber transmission. This integrated technique may be further extended to provide a powerful, system-defined particle selection and manipulation tool, with broad spectroscopic applications analogous to those demonstrated in free space via injection of particles onto the fiber surface[23] or selective deposition in a Paul trap.[24] The details and applications of this combined system are outlined below.

## 2. EXPERIMENTAL METHODS

### 2.1 Optical tweezers

To create the integrated tweezers-nanofiber system, we used a home-built tweezers based on Thorlabs model OTKB/M, with a 300 mW 1064 nm laser (hereafter referred to as the *tweezers* laser). It features an inverted microscope configuration, where the sample is placed over the oil-immersion objective (100x 1.25 NA) and illuminated from above. The tweezers includes a galvo-steered mirror pair in the incident beam path which allows modulation of the beam position in the focal plane for simultaneous trapping of multiple particles through time-sharing of the beam.

### 2.2 Optical nanofiber fabrication, mounting and integration

The most common means of nanofiber fabrication involves stripping, heating and stretching standard optical fiber, until its waist region reaches nanoscale dimensions. Tried and tested heat sources include $CO_2$ lasers,[25] ceramic microheaters,[26] and butane or hydrogen torches.[27] For basic taper production, a modest pulling rig can easily be assembled in the laboratory – the heat source (typically a few mm wide) remains stationary, while two linear stages draw the fiber outwards. This yields exponential taper shapes and high quality nanofibers with adiabatic transmissions close to 100%.[28] Introducing a lateral scanning function to the heat source immediately improves the rig functionality, potentially allowing the user to create and reproduce high quality nanofibers with arbitrary waist lengths, diameters and taper profiles.[27]

Using a hydrogen flame-brushed, heat-and-pull method[29], optical nanofibers with waist diameters of 530 nm were fabricated from Thorlabs 1060XP fiber. The nanofibers had taper and waist lengths of 29 mm and 2 mm, respectively, and measured transmissions greater than 90%. We selected a nanofiber diameter of 530 nm to satisfy the single-mode condition for 1064 nm[13] and ensure an enhanced evanescent field, while maintaining fiber robustness for handling. To facilitate nanofiber integration into the optical tweezers, a fiber mount was designed and 3D-printed to enable it to be stretched taut, tilted with respect to the horizontal plane, and moved vertically within the sample.

After fabrication, the tapered optical fiber was fixed to the mount with UV-curing optical glue, inspected under an optical microscope, and stretched until taut. Finally, the fiber mount was firmly attached to the tweezers' 3D stage, positioning the nanofiber centrally over the pre-mounted cover-slip (Fig. 1).

To facilitate stable co-planar focusing with the optical trap (1.25 NA objective with relatively small working distance), the nanofiber should be located within 100 μm of the sample cover slip surface due to the limited working distance of the objective. The tilt function of the mount allows the user to align the fiber parallel to the surface of the cover-slip; this is important for both even focusing of the fiber over the field of view, and to avoid contact between the fiber and the cover slip which result in complete loss of the fiber-guided light.

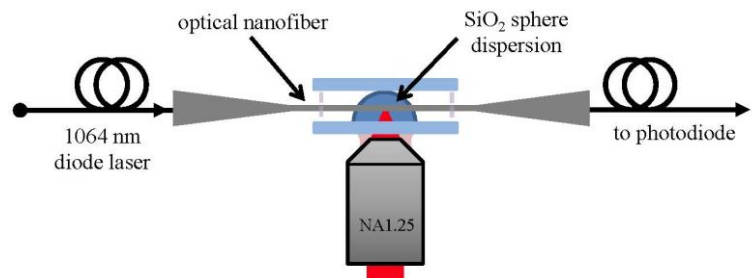

Figure 1. The nanofiber is integrated into the optical tweezers using a custom-designed mount enabling the fiber to be positioned carefully in the microparticle solution.

## 2.3 Nanofiber in dispersion

Silica microspheres of 2.01 µm, 3.13 µm and 5.06 µm (SS04N, SS05N, and SS06N Bangs Laboratories, Inc.) were diluted with ultrapure water (Nanopure™, Thermo Scientific) to concentrations of $10^5$-$10^6$ µl$^{-1}$. Prior to positioning the fiber, 100 µl of the dispersion was pipetted onto the center of the cover-slip (droplet ~ 12 mm diameter), and left for five minutes to allow the beads to settle to the bottom. Once mounted, one fiber pigtail was spliced to a 300 mW fiberized 1064 nm diode laser, hereinafter referred to as the *guided* laser, and the transmitted power was monitored at the fiber output. The nanofiber was horizontally adjusted and lowered until it touched the dispersion, at which point the fiber transmission was lost due to out coupling on contact with the surface. Adding a drop of water helps the fiber break the surface tension and restores transmission. The fiber was then lowered until it was just above the cover-slip (Fig. 1). Some oil was added to the objective, and the fiber was brought into focus. Finally, a second coverslip was positioned above the sample using short supports. This stabilizes the system by alleviating atmospheric air-flow across the droplet and serves to delay water evaporation, a cause of power fluctuations. With an exponentially-tapered fiber, the fiber transmission gradually increased until it reached a typical level of 80% near the cover-slip surface. The loss in this instance is repeatable, and largely due to scattering at the air-water interfaces of the thin fiber. However, if linearly-shaped tapers are used, the transmission is restored nearly to its initial value when the nanofiber is positioned close to the coverslip. This may be attributed to the larger fiber diameter as it enters and leaves the droplet in this geometry. Thus, the linear tapers prove to be more suitable for this type of applications due to a more uniform waist and increased robustness.

## 3. EXPERIMENTAL APPLICATIONS

### 3.1 Particle trapping and introduction to nanofiber

Before placing the fiber into the optical tweezers system, the 1064 nm guided light power was stable within 0.1%. Once placed in the dispersion, the guided laser power fluctuated by 0.3 to 0.5%. This increase in the fluctuations appears to arise from systemic noise in the combined tweezers-nanofiber system. In the first instance, single silica microparticles were trapped on the surface of the optical tweezers sample holder cover-slip. The sample stage was then lowered to bring the trapped particles into common focus with the nanofiber. Fig. 2(a) shows individually trapped particles in the vicinity of the nanofiber, with a 2.01 µm sphere in the top image and a 3.13 µm sphere in the bottom image. The trapped particle can be introduced into the evanescent field of the fiber by simply translating the sample stage, or by galvo-steering. Particles next to the nanofiber are shown in Fig. 2(b). The power of the guided laser through the optical nanofiber was monitored throughout such introductions and sharp drops in transmitted power were consistently observed due to particle presence. Fig. 2(c) shows typical fiber transmission plots recorded for both particle sizes, with markedly larger transmission drops recorded for the larger sphere at the same fiber position. Interestingly, these dips were not relative to the trapping power of the optical tweezers – we found that even at powers >100 mW there was negligible coupling of light from the trap to the nanofiber.

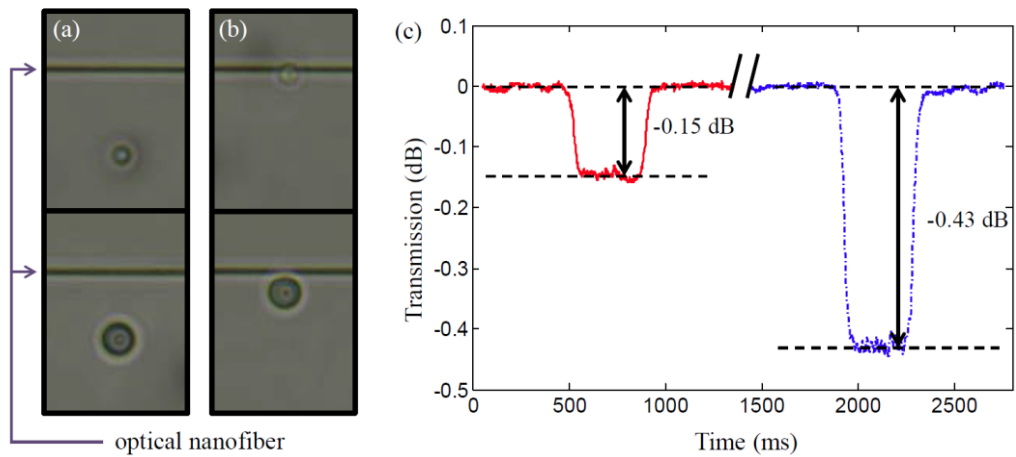

Figure 2. 2.01 µm (top) and 3.13 µm (bottom) silica microspheres trapped (a) near to and (b) in selective individual contact with the nanofiber surface. (c) Guided laser transmission dips due to particle scattering in the nanofiber evanescent field. Left (red solid) plot for 2.01 µm particle. Right (blue line-dot) plot for 3.13 µm particle.

Next, we considered the influence of nanofiber diameter on fiber transmission loss. Similar transmission effects to those in Fig. 2(c) were systematically recorded for 2.01 μm, 3.13 μm and 5.06 μm particles at 100 μm interaction intervals along the fiber – starting in the tapered region, continuing across the nanofiber waist, and finally along the opposite taper. The fiber profile was then measured via scanning electron microscopy (SEM) (Fig. 3(a)) and the results correlated (see Fig. 3(b)). The decreasing nanofiber diameter induces dramatic and increasing scattering in the particle; this scattering also increases with particle size. This effect can be associated with varying penetration depths and intensity profiles of the evanescent fields at different fiber diameters for a given wavelength; experimental knowledge of this relationship is important for emerging techniques for site-specific particle detection in low-density solutions.

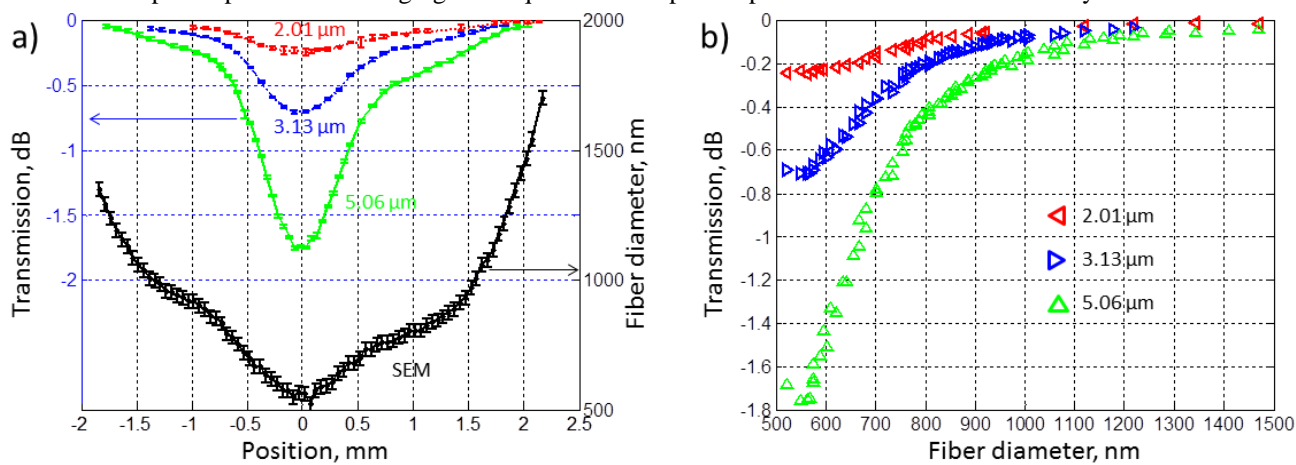

Figure 3. a) Fiber transmission when in contact with a 2.01 μm (red line), 3.13 μm (blue line), and 5.06 μm (green line) silica sphere, and the fiber diameter as measured from SEM images (black line), as a function of position along the fiber axis. b) Fiber transmission as a function of fiber diameter when probed by silica particles with diameters of 2.01 μm (red left-pointing triangles), 3.13 μm (blue right-pointing triangles) and 5.06 μm (green up-pointing triangles).

### 3.2 Particle jet

For the aforementioned applications, the particles were trapped in the tweezers using powers of mW to tens of mW, depending on particle size and the number of traps. Typically, the trapping threshold for 2-3 μm spheres was 1 mW per particle, but higher powers were used to make the traps stiffer and to stabilize the particle position. With increasing power, more particles move towards the tweezers' beam focus, occupying the trap collectively or expelling previously trapped particles. This can, in many circumstances, be more of a hindrance to experiments than a benefit. However, such a trap can create an imbalanced particle 'jet' whereby particles in close proximity to each other are continuously 'pulled in' towards the bottom of the focus and expelled at the top. If placed close to a nanofiber, such a jet can be used to source a continuous 'conveyor belt' of particles, which are propelled along the fiber (Vid. 1). This simple yet efficient system has applications in particle sorting, particularly for particles already located on a surface within a sample, as opposed to being injected from an external source via microfluidic techniques.

### 4. CONCLUSION

We have presented a combined optical nanofiber-optical tweezers integrated platform for particle propulsion and manipulation. This system is a promising tool for studying evanescent field-particle interactions. The measured dependence of fiber transmission as a function of both to sphere and nanofiber diameter, has broad applications in system calibration and multiparticle size-sorting.

The tweezers and nanofiber intensities can be modulated to trap at a specific surface position on the fiber, or propel in both directions by balancing bidirectional nanofiber fields. This setup is also perfectly placed for trapping multiple particles, and was successfully used for studying optical binding interactions occurring in the evanescent field of an optical nanofiber[22]. Additionally, the particle jet-to-nanofiber presented has potential in high-throughput cell sorting or cytometry, site clearing, and particle accumulation.

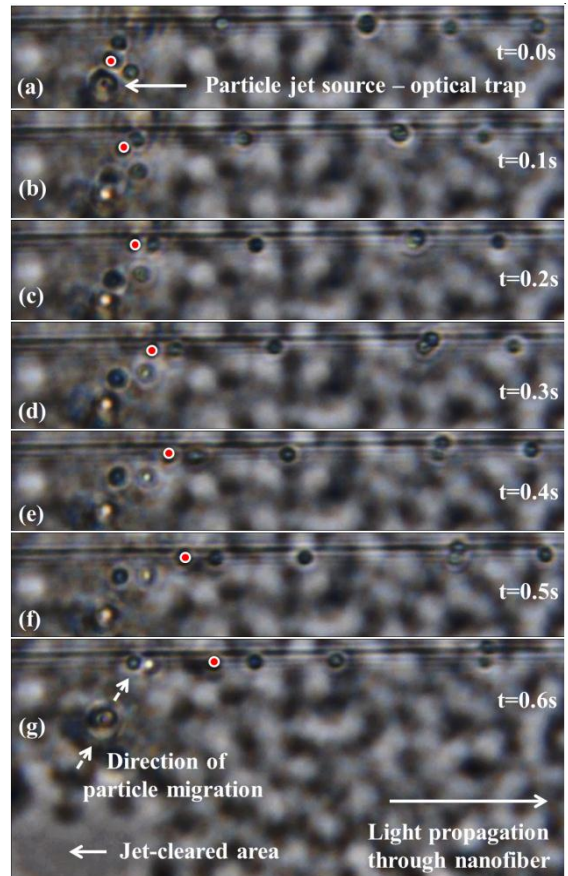

Video 1. (a) Particle 'jet', feeding a continuous stream of 2.01 μm silica spheres to the nanofiber. The particles are then propelled along the surface of the nanofiber by the evanescent field (b-g). http://dx.doi.org/doi.number.goes.here

Although the experiments described above utilized silica microspheres, the techniques are also applicable to optical interaction studies and manipulation of other dielectric or biological specimens, notably polystyrene spheres and living cells. Extended functionality of the tweezers will introduce further experimental possibilities – for example, introducing high precision calibrated particle tracking via a quadruple photodiode[30] or high-speed camera[31] would enable site-specific force measurements of the evanescent field mediated particle-particle and fiber-particle interactions.

Finally, the nanofiber field may be tailored by changing the input wavelength, power and light polarization; the combined tweezers-nanofiber system thus has further potential for use in fluorescence and spectroscopy studies, using the nanofiber as a passive (collection) or active (excitation) probe. The evanescent field structures of non-uniform fiber elements,[32] and complex evanescent fields under counter propagating light[18] and higher mode propagation[33,34] may also be investigated using this system.